\documentclass[11pt]{aastex}
\usepackage{graphicx,emulateapj5,apjfonts}
\usepackage{onecolfloat}
\usepackage{epsf}

\shortauthors{Bell et al.}
\shorttitle{The merger rate of massive galaxies}

\newcommand{\peg}{{\sc P\'egase }}
\newcommand{\combo}{{COMBO-17} }

\slugcomment{{\sc To appear in ApJ:}Dec 1 2006 }

\begin{document}


\def\head{

\title{The merger rate of massive galaxies }

\author{Eric F.\ Bell$^1$, Stefanie Phleps$^{2,3}$,
Rachel S.\ Somerville$^1$, Christian Wolf$^4$, 
Andrea Borch$^5$, and Klaus Meisenheimer$^1$}
\affil{$^1$ Max-Planck-Institut f\"ur Astronomie,
K\"onigstuhl 17, D-69117 Heidelberg, Germany; \texttt{bell@mpia.de}\\
$^2$ Institute for Astronomy, University of Edinburgh, Royal Observatory,
	Blackford Hill, Edinburgh EH9 3HJ, UK \\
$^3$ Max-Planck-Institut f\"ur extraterrestrische Physik,
Giessenbachstra{\ss}e, D-85748 Garching, Germany \\
$^4$ Department of Physics, Denys Wilkinson Bldg., University
of Oxford, Keble Road, Oxford, OX1 3RH, UK \\
$^5$ Astronomisches Rechen-Institut, M\"onchhofstr. 12-14, D69120 Heidelberg,
Germany \\
}

\begin{abstract}
We calculate the projected two point correlation function 
for samples of luminous and massive galaxies in the COMBO-17
photometric redshift survey, focusing particularly on 
the amplitude of the correlation function at small 
projected radii and exploring the constraints
such measurements can place on the galaxy merger rate. 
For nearly volume-limited samples with $0.4<z<0.8$, we find  
that 4$\pm$1\% of luminous $M_B < -20$ galaxies are in
close physical pairs
(with real space separation of $< 30$\,proper kpc).
The corresponding fraction for massive galaxies with 
$M_* > 2.5 \times 10^{10} M_{\sun}$ is 5$\pm$1\%.
Incorporating close pair fractions from the literature, the 
2dFGRS and the SDSS, we find a fairly rapid evolution of the merger
fraction of massive galaxies between 
$z = 0.8$ and the present day.  Assuming that the major merger
timescale is of order the dynamical timescale for close massive
galaxy pairs, we tentatively infer that $\sim 50$\% (70\%) 
of all galaxies with present-day masses
$M_* > 5 \times 10^{10} M_{\sun}$
(remnants of mergers between 
galaxies with $M_* > 2.5 \times 10^{10} M_{\sun}$) 
have undergone a major merger since $z = 0.8\,(1)$: 
major mergers between massive galaxies are a significant
driver of galaxy evolution over the last eight billion years.
\end{abstract}

\keywords{galaxies: general ---  galaxies: interactions ---
galaxies: evolution}
}

\twocolumn[\head]

\section{Introduction}

Galaxy mergers are ubiquitous in a hierarchical universe, and 
are predicted to be an important mode of galaxy growth, particularly
at early times in cosmic history \citep[e.g.,][]{keres03,maller05}. 
Mergers may be an important feature of the growth of 
massive early-type galaxies \citep[e.g.,][]{toomre72,barnes96,khochfar03,
vandokkum05,naab06,bell06}.  It is possible that 
mergers may also be an important driver of disk galaxy evolution:
gas-rich mergers may 
conserve enough angular momentum
to form a disk \citep{robertson04}.
The non-circular gas motions
induced by the rapidly-changing potential may drive gas inflows, 
igniting intense star formation \citep[see, e.g.,][for a review]{sanders96}, 
feeding pre-existing supermassive black holes, 
enhancing AGN activity, and perhaps
even driving a galaxy-scale superwind, evacuating the galaxy of gas
\citep{springel05,dimatteo05}.  

Despite their importance, it has proven challenging to measure
the rate of major galaxy merging, and its evolution with cosmic epoch.
The measurement of the galaxy interaction rate by counting the incidence
of strongly-disturbed galaxies (with strong asymmetries, double nuclei, 
or prominent tidal tails) has provided important constraints on merger
rate \citep[e.g.,][]{lefevre00,conselice03,conselice05,lotz06}, but suffers
from uncertainties: minor gas-rich interactions may produce much 
more spectacular results than a major merger between two 
spheroid-dominated galaxies, and the timescales over which 
merger signatures are visible is highly dependent on 
orbits, gas content, and mass ratio.
Another powerful method for exploring the galaxy merger rate is 
to measure the incidence of close pairs of galaxies: it allows access to 
the properties of the progenitors (and therefore, e.g., stellar
mass ratio of the merger), is straightforward to quantitatively
measure, and can be modeled using current generations of galaxy
formation models.  Yet, this has proven to be a reasonably
challenging endeavor: contamination by projection, luminosity
boosts by interaction-induced star formation, and small number
statistics are significant challenges and are not easily circumvented.

In this paper, we present our first attempt at addressing 
this issue using the COMBO-17 photometric redshift survey (\S \ref{data}).
This analysis uses a large sample, attempts to correct
the close pair fraction estimates for projection, and
incorporates stellar mass estimates, making it highly
complementary to other important recent attempts
at measuring the evolution of close pair fraction
\citep[e.g.,][]{patton02,lin04}.
We split the problem into two aspects.
Firstly, we measure the fraction of galaxies in 
$r < 30$\,kpc separation pairs (in real space) through 
analysis of the projected correlation function of galaxies from 
COMBO-17 (\S \ref{method}).  This is a well-understood and well-posed problem, 
with a clear and well-constrained outcome (\S \ref{results}).  
We subsequently use estimates of merger timescale
to explore implications for the merger rate of galaxies: inasmuch
as this part of the analysis makes use of somewhat uncertain
merger timescales and assumes that all real space close
pairs will merge, this part of the analysis is much less 
robust (\S \ref{disc}).    
Throughout, we assume $\Omega_{\rm m} = 0.3$, $\Omega_{\rm
m}+\Omega_{\Lambda} = 1$, and $H_0 = 70 $\,km\,s$^{-1}$\,Mpc$^{-1}$.

\section{The Data} \label{data}

To date, \combo has surveyed three disjoint
$\sim 34' \times 33'$ southern and equatorial fields
to deep limits in 5 broad and 12 medium passbands.
Using these deep data in conjunction with
non-evolving galaxy, star, and AGN template spectra, objects
are classified and redshifts assigned for $\sim 99$\% of the
objects to a limit of $m_R \sim 23.5$.  Typical galaxy redshift accuracy
is $\delta z/(1+z) \sim 0.02$ \citep{wolf04},
allowing construction of $\sim 0.1$ mag accurate
rest-frame colors and absolute magnitudes (accounting for distance
and $k$-correction uncertainties).  Astrometric accuracy is
$\sim 0.1$\arcsec.  Owing to reduced depth close to the edges of 
the fields, we discard galaxies $<1'$ from the image edge.

As we are concerned with the clustering of galaxies on 
small angular scales, we have tested how the detection of 
galaxies by COMBO-17 is affected by having a nearby luminous neighbor.
We extracted the images of 400 isolated massive galaxies (galaxies which 
were included in our sample)
in the COMBO-17 image of the Extended Chandra Deep South, and placed
these postage stamps of real massive galaxies with a distance 
$0'' < r < 10''$ from another massive galaxy.  We then put this modified
image through the COMBO-17 object detection pipeline, allowing 
us to determine the fraction of these inserted massive galaxies
which were recovered by COMBO-17's pipeline as a function 
of distance from the primary galaxy.  We found that the 
detection fraction was independent of distance at $r > 2''$; only
at $r < 2''$ (corresponding to $\sim 15$\,kpc at the redshift of interest) 
was there evidence for substantial incompleteness 
in object recovery.  Accordingly, we do not use information 
from pairs with separations $<15$\,kpc in what follows.

\citet{borch} estimated the stellar mass of galaxies in
\combo using the 17-passband photometry in conjunction with
a non-evolving template library derived using the
\peg stellar population model \citep[see][for a description
of an earlier version of the model]{fioc97}.  The masses were
derived using a \citet{kroupa93} stellar IMF; the use of a
\citet{kroupa01} or \citet{chabrier03} IMF would have yielded
similar stellar masses, to within $\sim 10$\%.  The redder
templates have smoothly-varying exponentially-declining 
star formation episodes and a low-level constant
star formation rate; the bluer templates have a recent burst
of star formation superimposed (thus, {\it ongoing} tidally-induced
bursts of star formation are approximately
accounted for).  Such masses
are quantitatively consistent with those derived using a simple
color-stellar M/L relation \citep{bell03}, and comparison of
stellar and dynamical masses for a few $z \sim 1$ early-type
galaxies yielded consistent results to within their
combined errors \citep[see][for more details]{borch}.
Random stellar mass errors are $< 0.3$\,dex on a galaxy-by-galaxy
basis, and systematic errors in the stellar
masses (setting the overall mass scale and its redshift evolution)
were argued to be at the 0.1\,dex level.

\section{The Method} \label{method}

As stated in the introduction, our goal is to estimate
as accurately as possible the fraction of galaxies which 
have a companion satisfying our selection criteria within 
a certain physical distance. 
The accuracy of photometric redshifts is clearly 
insufficient to directly estimate the redshift
space correlation function of galaxies directly: 
COMBO-17's typical redshift error translates into 
$\sim 200$\,Mpc along the line-of-sight.  Consequently,
we separate this problem into two parts: construction of a
projected correlation function, and subsequent de-projection of
this correlation function into a real-space correlation function.

\subsection{Estimating the projected correlation function}

As a first step, we calculate the projected two-point correlation 
function of galaxies \citep{davis83}.  The projected correlation function 
$w(r_p)$ is the integral of the real space correlation function
$\xi (r)$ along the line of sight: 
\begin{equation}
w(r_p) = \int^{\infty}_{- \infty} \xi ( [r_p^2 + \pi^2]^{1/2}) d \pi,
\end{equation}
where $r_p$ is the transverse distance between two galaxies and $\pi$ is
their line-of-sight separation.  It is clear that traditional close
pair counts are an integral over small projected radii of the 
projected two-point correlation function
for galaxies with particular properties, multiplied by the space
density of such galaxies.

In practice, we estimate $w(r_p)$ for various samples of 
galaxies using the following scheme.  
We construct a histogram of the number
of galaxy pairs with given properties (as defined later
in \S \ref{results}) and $|\Delta z|<0.05$ 
as a function of projected physical 
separation, and for a randomly-distributed
mock galaxy sample.  
The mock samples were generated from the real data by bootstrapping
the data multiple times, assigning random positions and 
fields, and applying a small Gaussian redshift offset ($\sigma_z = 0.04$).
Masks defining the field edges and the areas around bright
stars were applied in the same manner to the data and random fields.
Projected auto-correlations $w(r_p)$ were estimated
from these histograms by constructing the ratio 
$w(r_p) = \Delta ({\rm DD/RR - 1})$, where
$\Delta$ is the path length being integrated over, DD is 
the histogram of separations between real galaxies and RR is the
histogram of separations of mock catalog galaxies.   Other formulations 
($\frac{\rm DD-2DR-RR}{\rm RR}$ and ${\rm DD/DR - 1}$, where DR
is the histogram of separations of real and mock galaxies) were verified to be 
equivalent to within the errors.  

One will see that, in order to preserve S/N, 
we did not integrate along
the entire line of sight in calculating $w(r_p)$, 
rather along $\pm 0.05$ in redshift path length
from the galaxy of interest.  
Therefore, we must correct our estimate of $w(r_p)$ 
to account for pairs missed because their photometric
redshift errors took them erroneously out of the redshift
range being integrated over.  
Extensive comparison with spectroscopic
redshifts \citep{wolf04} 
has shown that a Gaussian with the following $R$-band apparent 
magnitude-dependent width is an adequate representation of 
the photometric redshift errors:
$\sigma_z \sim 0.01134 [1 + 10^{0.8(m_R-21.5)}]^{1/2}$.
The average $\sigma_z$ for the galaxy sample of interest
was evaluated, and the redshift difference distribution of galaxy 
pairs was taken to be described by a Gaussian with width
$\sigma_{\rm pair} = \sqrt{2} \sigma_z$.  Accordingly, the fraction of genuine 
galaxy pairs included in the sample is: 
\begin{equation}
f = \int_{-0.05}^{0.05} \frac{1}{\sqrt{2 \pi} \sigma_{\rm pair}} e^{-z^2/2\sigma_{\rm pair}^2} dz,
\end{equation}  (i.e. the fraction 
missed was $1-f$).  The 
estimate of $w(r_p)$ was then multiplied by $1/f$ to account for the missing
pairs, a correction 
of $\la 30$\% in all cases.  No correction for the 
integral constraint was applied: \citet{phleps05}
show in their \S 4.4 that such corrections are $< 0.5$ Mpc in $w(r_p)$ for 
$0.4<z \le 0.8$ luminous/massive galaxies in COMBO-17,
which given our focus on the strongly correlated smallest scales
is a $\ll 1$\% correction. 

\subsection{Estimating the real space correlation function}

Assuming that the real-space and projected correlation functions
can be adequately fit with power laws, the parameters of the 
two fits are intimately related.  If $\xi(r) = (r/r_0)^{-\gamma}$, 
then $w(r_p) = C r_0^{\gamma} r_p^{1-\gamma}$, where
$C = \sqrt{\pi} \frac{\Gamma ([\gamma - 1]/2)}{\Gamma (\gamma/2)}.$
Thus, we have adopted the approach of fitting the $w(r_p)$ 
estimated from the data as a function of projected radius, then using the 
above relations to estimate $r_0$ and $\gamma$.
It is also possible to directly
estimate $\xi(r)$ from $w(r_p)$ using the Abel integral 
\citep{davis83}: we choose not to adopt this method 
by default in this work because the direct inversion 
is rather noise-sensitive, and because 
we are interested in very close pairs $r < 30$\,kpc, and 
some power-law extrapolation of $\xi(r)$ would be necessary at any rate
to fill in the closest $< 15$\,kpc separations (see above). 
We have confirmed that in the 
best-posed cases where the $w(r_p)$ are well-measured 
that {\it i)} the correlation functions are well-parameterized
by a power-law, and {\it ii)} inverting $w(r_p)$ directly
into $\xi(r)$ gives very similar answers to the power-law
parameterization. 

\subsection{Estimating the close pair fraction}

Given a good parameterization of the real space correlation 
function of galaxies on small $\la 30$\,kpc scales, it 
is straightforward to define a real space close pair 
fraction \citep[as discussed by][]{patton00,masjedi06}.  
Recall the definition of the real space correlation 
function: $\delta P = n [1+\xi(r)] \delta V$, where $\delta P$ is 
the probability of a secondary galaxy occupying a volume $\delta V$ a distance
$r$ from the galaxy of interest, and $n$ is the space
density of the secondary galaxies (equal to the space density of 
the primary galaxies for an autocorrelation).  Therefore, the probability
of a galaxy being within a distance $r_f$ of another galaxy 
satisfying our selection criteria is:
\begin{eqnarray} 
P(r<r_f) &  = &  \int^{r_f}_0 n [1+\xi(r)] dV, \\
         &  \approx & 4 \pi n \int_0^{r_f} r^2 \xi(r) dr, 
\end{eqnarray}
given that $\xi(r) \gg 1$ at all radii of interest for this paper.
Therefore, parameterizing the real-space correlation function
as $\xi(r) = (r/r_0)^{- \gamma}$, one obtains:
\begin{eqnarray}
P(r<r_f) = \frac{4 \pi n}{3 - \gamma} r_0^{\gamma} r_f^{3-\gamma}.
\end{eqnarray}
It is worth noting that because typically $\gamma \sim 2$, 
$P(r<r_f) \propto r_f$ to first order: i.e., that there is 
roughly an equal contribution of galaxies in each radius
bin to the total close pair fraction. 

\section{Results} \label{results}

\begin{figure}[t]
\begin{center}
\epsfxsize 9.0cm
\epsfbox{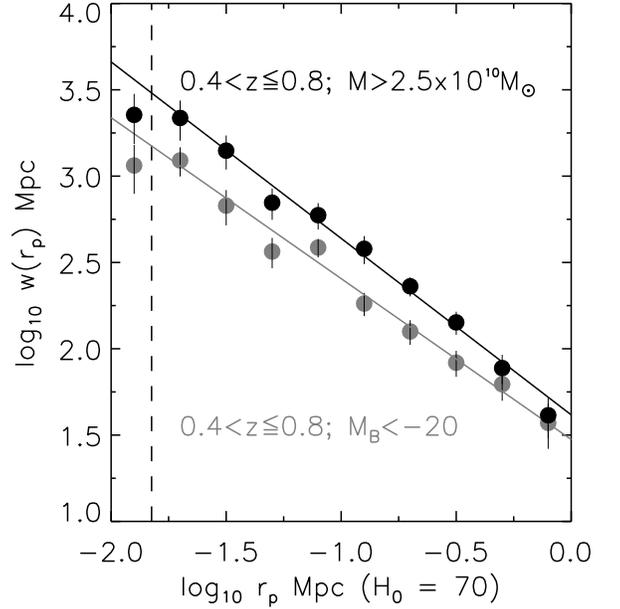}
\end{center}
\caption{\label{fig:cf}  
{\it Grey:} Projected correlation function $w(r_p)$ for
the $0.4<z<0.8$ volume-limited sample of 
luminous $M_B < -20$ galaxies.  The power
law fit to $w(r_p)$ is overplotted, and parameters given in Table
\ref{tab}.  The vertical dashed line at 15\,kpc shows the radius within
which COMBO-17's object detection pipeline no longer reliably separates
nearly equal-luminosity close galaxy pairs: this corresponds to 
the radius at which the correlation function starts to deviate strongly
from a power law.
{\it Black:} The projected correlation 
function of the volume-limited $0.4<z<0.8$ sample of
massive $M_* > 2.5 \times 10^{10} M_{\sun}$ galaxies.  
}
\end{figure}

\begin{table}
\begin{center}
\caption{Real space correlation function parameters and 
         close pair fractions {\label{tab}}}
\begin{tabular}{lcccc}
\hline
\hline
Sample & $r_0$/Mpc & $\gamma$ & $n$/Mpc$^3$ & $P(r<r_f)$ \\
\hline
\multicolumn{5}{c}{Auto-correlations} \\
\hline
(1) & 3.1$\pm$0.5 & 1.93$\pm$0.08 & 0.0162 & 0.04$\pm$0.01 \\
(2) & 3.6$\pm$0.5 & $2.02 \pm 0.07$ & 0.0091 & 0.05$\pm$0.01 \\
\hline
\multicolumn{5}{c}{Cross-correlations} \\
\hline
(3)& $3.2 \pm 0.5$ & $1.89 \pm 0.09$ & 0.0136 & $0.028 \pm 0.005$ \\
(4) & $3.7 \pm 0.5$ & $1.98 \pm 0.08$ & 0.0061 & $0.029 \pm 0.005$ \\
\hline \\
\end{tabular}
\end{center}
Sample (1) is a luminosity and volume-limited sample with $M_B < -20$ and $0.4 < z \le 0.8$ \\
Sample (2) is a stellar mass and volume-limited sample with $M_* > 2.5 \times 10^{10} M_{\sun}$ and $0.4 < z \le 0.8$ \\
Sample (3) has primary galaxies with $M_B < -20.3$ and $0.4 < z \le 0.8$; secondary galaxies must be 1.2 mag fainter or less than their primary galaxy \\
Sample (4) has primary galaxies with $M_* > 3 \times 10^{10} M_{\sun}$; 
secondary galaxies must be 1/3 or more of their primary galaxy's mass to be counted as a pair \\
For samples (3) and (4), the space density of secondary galaxies is 
used to determine the pair fraction \\
\end{table}

As stated earlier, our goal is to understand the role of merging in
driving the evolution of massive galaxies.  Therefore, we study the
close pair fraction of galaxies selected in two ways: luminous $M_B <
-20$ galaxies, and massive $M_* > 2.5\times 10^{10} M_{\sun}$
galaxies.  We restrict the sample to galaxies in the redshift interval
$0.4<z<0.8$.  Each sample is $\sim 99$\% complete across the entire
redshift range of interest, forming a nearly volume-limited sample.
The auto-correlation function $w(r_p)$ for each sample has been
derived using the methods outlined above.  A power law is fit to the
$w(r_p)$ values for the whole sample, and the close pair fraction
$P(r<r_f)$ derived from this power law fit coupled with the measured
number density of galaxies meeting our selection criteria.  Error bars
in all quantities were derived for each field separately, adopting
Monte Carlo errors of 0.1 dex in luminosity/mass and accounting for
counting uncertainties in the histogram of real galaxy pairs DD.  
These were then combined in quadrature
and divided by $\sqrt{(N_{\rm field}-1)}=\sqrt{2}$.  It is important
to note for this paper we have used proper coordinates to calculate
the correlation functions and space densities.

\begin{figure}[th]
\begin{center}
\hspace{-1.cm}
\epsfxsize 8.5cm
\epsfbox{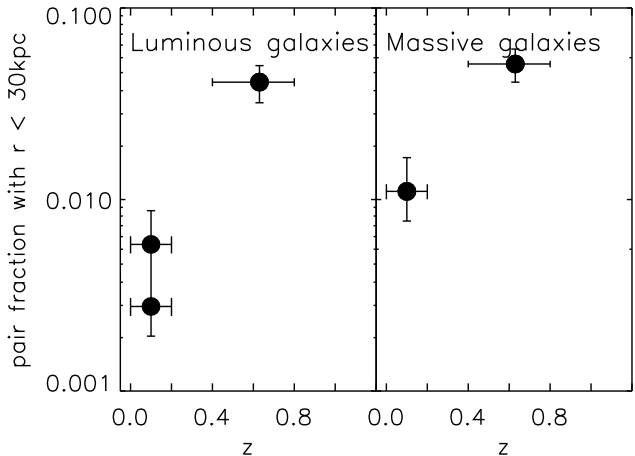}
\end{center}
\caption{\label{fig:frac} 
The close pair fraction (with real space separations 
$< 30$\,kpc) of luminous ($M_B < -20$; left panel) 
and massive ($M_* > 2.5 \times 10^{10} M_{\sun}$; right panel)
galaxies.  The $z \sim 0.6$ data points are from the present work, 
and the $z \sim 0.1$ data points are estimated (roughly) from 
the 2dFGRS (left) and SDSS (right).  In the case of the 2dFGRS, two
estimates are shown: the lower estimate is for $M_B < -20$ galaxies, 
while the upper estimate is for galaxies with $M_B < -19.4$ --- the 
plausible descendants of $M_B < -20$ galaxies at $z = 0.6$.
}
\end{figure}

The results are shown in Figs.\ \ref{fig:cf} and \ref{fig:frac}.
One can see that the power-law
parameterization to $w(r_p)$ is an acceptable description of the 
data on sub-Mpc scales, and that $w(r_p)$ is measured with interesting
accuracy even to $\sim 20$\,kpc scales (allowing reasonably robust 
measurement of $P(r<r_f)$); we showed earlier that object
detection failed for separations $< 15$\,kpc, therefore
power law fits to $w(r_p)$ are determined only 
for $15 < r/{\rm kpc} < 1000$, and are extrapolated
inwards.   The resulting close pair 
fractions $P(r<r_f)$, adopting $r_f = 30$\,proper kpc, are
shown in Fig.\ \ref{fig:frac}.

Many previous studies have attempted to estimate the major 
merger rate, i.e., the rate of galaxies merging with galaxies
with mass ratios between 1:1 and 3:1.  Accordingly, we have
also estimated the cross-correlations between galaxies in our sample
and potential major merger partners: for the luminous galaxy samples,
between $M_B < -20.3$ galaxies and galaxies between $0$ and $1.2$\,mag
fainter than the primary galaxy; and for the massive galaxy samples
between $M_* > 3\times 10^{10} M_{\sun}$ galaxies and galaxies
with between $1/3$ and the same mass as the primary.  The samples
are close to volume limited in each case; the faintest 
red sequence galaxies are missing at $z \ga 0.6$ in the 
secondaries (the blue cloud secondary galaxies are complete
at all redshifts), leading to a $\la 7$\% incompleteness in
the secondary sample; we do not correct for this 
incompleteness.  The number density of secondary galaxies is 
not trivial to calculate as each primary galaxy has a different
set of secondary galaxies.  We 
estimated the `characteristic' number density of secondary galaxies
by evaluating the number density of potential secondary galaxies
for each primary galaxy, and then averaging these number densities.
The cross-correlation parameters are very similar to those for 
the autocorrelations, with
the modest differences between the values of  $P(r<r_f)$
for the auto- and cross-correlations driven by the differences in 
the number densities.  We give 
parameters of power-law fits to the $w(r_p)$ estimates in 
Table \ref{tab}.

\section{Discussion} \label{disc}

\subsection{Comparison with local pair fractions derived using the same method}

Using published studies, one can use the methodology presented
in this paper to estimate the local pair fraction. 
For luminous galaxies, we use results from the Two degree Field Galaxy
Redshift Survey (2dFGRS; Colless et al.\ 2001).
\citet{norberg02_two} presented the two-point correlation 
function for galaxies with $M_{b_J} - 5 \log_{10}h_{100} < -19.5$ 
(corresponding to $M_B \la -20.1$ adopting $b_J \sim B - 0.15$ from 
Norberg et al.\ and converting to $h_{100} = 0.7$).  The 
values of $r_0$ and $\gamma$ are $3.5\pm 0.4$\,Mpc and $1.8 \pm 0.1$ 
respectively, and 
were defined only outside 
100$h^{-1}_{100}$\,kpc; the inwards extrapolation required
for this analysis is unconstrained and therefore this estimate
should be regarded with due caution.  Adopting 
the luminosity function from \citet{norberg02_lf}, a density
of galaxies with $M_B < -20$ of 0.0019(2)$\,$Mpc$^{-3}$ was derived, 
giving a pair fraction estimate of 0.003(1). 
An unavoidable complication is that of the fading of galaxies towards
lower redshift  as their stellar populations fade; thus, a sample of 
somewhat fainter galaxies at lower redshift may be a better 
conceptual match to  
the population of distant $M_B < -20$ galaxies.  Adopting a somewhat
fainter cut of $M_{B(z=0)} < -19.4$ (corresponding to 1 mag fading per 
unit redshift in the rest-frame B-band), one derives instead a largely
unchanged clustering signal \citep{norberg02_two}, a larger density
of 0.0039(4)$\,$Mpc$^{-3}$, and a larger pair fraction of 0.006(2).
Two points become clear: two-point correlation function parameters 
are substantially less sensitive to limiting depth than the number density
or close pair fraction; and, the fading of stellar populations 
is a considerable
complicating factor in studying the evolution of close pair statistics
for luminosity-selected samples.

Stellar mass-limited samples largely overcome the last of these 
two challenges.
A large number of galaxies in the main galaxy sample of the 
Sloan Digital Sky Survey \citep[SDSS;][]{york00} have estimates of their
stellar mass.  We use the 
two-point correlation function
of mass-limited samples \citep{li05} and the space density
of massive galaxies \citep{bell03}.  Rough fits to Li et al.'s 
$10.2 < \log M_*/M_{\sun} < 10.7$
and $10.7 < \log M_*/M_{\sun} < 11.2$ samples 
gave correlation function parameters $r_0 \sim 4.2 \pm 0.4$ and 
$\gamma \sim 1.85\pm 0.05$: given a number density of 
$M_* > 2.5 \times 10^{10} M_{\sun}$ galaxies of 0.0040(4) galaxies/Mpc$^3$
from the SDSS stellar mass function of \citet{bell03}, we estimate
a $P(r<30\,{\rm kpc}) \simeq 0.011(5)$.  
Again, the correlation function was defined only 
outside 100$h^{-1}_{100}$\,kpc; for both the SDSS and the 2dFGRS a direct
redetermination of the pair fractions from the data would be 
preferred to these rough estimates.  The average redshift of the 
galaxy samples in both cases is $\langle z \rangle \sim 0.1$.

These estimates are included in Fig.\ \ref{fig:frac} as the $z \sim 0.1$
estimates.
It is worth noting in both cases that there is clear evidence for a 
dramatically-reduced fraction of galaxies in close ($\le 30$\,kpc separation) 
physical pairs at the present day, compared to the pair fraction 
at $z \sim 0.6$.  This will be discussed in more detail later.

\subsection{The relationship between true
close pairs and projected close pairs}  \label{proj}

In order to compare our measurements to others in the literature, 
it is necessary to explore the relationship
between the fraction of galaxies in real space close pairs
and the projected close pair galaxy fraction.  The projected
space close pair fraction is the real space close pair
fraction plus a contribution from more distant galaxies
along the line of sight\footnote{It is worth remembering that
the projection of very distant fore- and background 
galaxies has been automatically removed
when calculating $w(r_p)$.}.  Specifically, the projected
close pair fraction:
\begin{equation}
P'(r_p < r_f) = P(r<r_f) + \int_{r_f}^{\infty} 4 \pi r^2 n[1+\xi(r)] [2/\pi \sin^{-1}(r_f/r)]^2 dr,
\end{equation}
following previous notation.  In our particular case, $r_f = 30$\,kpc, 
and using the observed luminous/massive galaxy correlation 
functions and number densities as input, we find that 
65\% of projected close luminous pairs have real space
separations $< 30$\,kpc; the corresponding fraction 
for massive galaxies is 69\%.  It is important to note
that this `contamination' is with galaxies which are 
correlated with the host (i.e., those which are primarily
nearby with $30 < r/{\rm kpc} \la 1000$), and would likely
have very similar redshifts to the primary galaxy (i.e., much of
contamination is suffered by spectroscopic close
pair samples).  It is important to note that the 
exact fraction depends on the detailed form of the correlation 
function and should not be blindly adopted by workers using 
rather different sample cuts: 
in particular, our estimate is slightly
higher than the estimate of $\sim 50$\% from 
\citet{patton00}, which was derived using a very similar approach
with a less clustered lower-luminosity parent sample.

\subsection{Comparison with published merger fraction determinations}
\label{comp}

\begin{figure}[th]
\begin{center}
\hspace{-1.cm}
\epsfxsize 8.5cm
\epsfbox{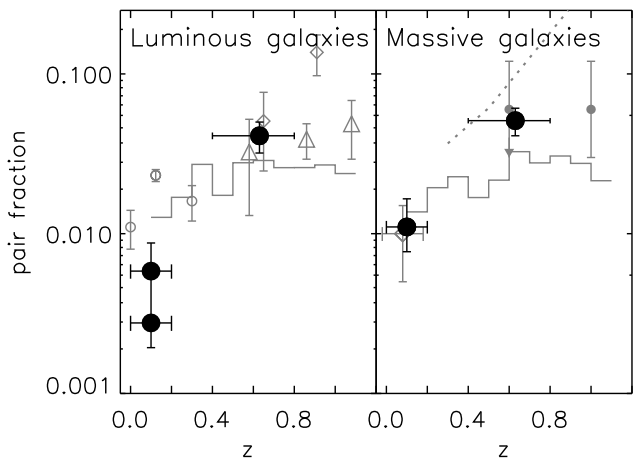}
\end{center}
\caption{\label{fig:frac2} 
The close pair fraction (with real space separations 
$< 30$\,kpc) of luminous ($M_B < -20$; left panel) 
and massive ($M_* > 2.5 \times 10^{10} M_{\sun}$; right panel)
galaxies.  {\it Left:} Gray data points show the 
merger fraction of galaxies with
$M_B \la -20$, taken from a variety of sources (see \S \ref{comp} 
for details).  
The Somerville et al.\ 
model prediction of the major merger fraction for $M_B < -20$
galaxies is shown
as a solid gray line for the redshift range $0.1<z<1.1$.
The $z \sim 0.1$ data point is derived from the 2dFGRS.
{\it Right:} Gray solid circles show the merger
fraction of galaxies with $M_* > 10^{10} M_{\sun}$ from 
\citet{conselice03}, and the Somerville et al.\ 
model prediction of the major merger rate 
of massive galaxies is shown as a solid gray line.
The merger fraction predicted by \protect\citet{maller05}
is shown by the grey dotted line.
The $z \sim 0.1$ black data point is derived from the SDSS, and
the gray diamond is taken from \protect\citet{xu04}.  
}
\end{figure}

In Fig.\ \ref{fig:frac2}, we have attempted to show close
pair or merger fractions from a variety of published works, 
to compare with our determinations.  Such an exercise is 
not trivial, as very different methods were
used in many cases.

\subsubsection{The $M_B < -20$ sample} 

The vast majority of previous determinations of close pair
fractions have been derived for luminous $M_B \la -20$ 
galaxies.  At $z < 0.3$, we show results 
for \citet{patton02} and \citet{depropris} 
as open circles.  \citet{patton02} adopts a radius range
$5 h^{-1}_{100} \le r \le 20 h^{-1}_{100}$\,kpc; adopting 
$H_0 = 70\,$km\,s$^{-1}$\,Mpc$^{-1}$ this becomes 
$7 \le r \le 30\,$kpc (i.e., some 1/4 of the galaxy pairs are not counted).  
Patton et al.'s default magnitude
range is $-21 \le M_B - 5\log_{10} h_{100} \le -18$; corresponding to 
$-22 \la M_B \la -19$ for $H_0 = 70\,$km\,s$^{-1}$\,Mpc$^{-1}$.
Using their table 3, one can convert their results to a narrower range
in absolute magnitude $-21 \le M_B - 5\log_{10} h_{100} \le -19$, or
$-22 \la M_B \la -20$ in our units, by dividing by 2.5.   
Accordingly, we adjust Patton et al.'s values upwards by 4/3 (to account
for $r < 7\,$kpc pairs) and downwards by a factor of 2.5 (to account for
the magnitude range).  \citet{depropris} adopt a magnitude 
range of $-22 \le M_B - 5\log_{10} h_{100} \le -19$, corresponding
to $-23 \la M_B \la -20$ for our choice of $H_0$, thus the only 
correction applied is 4/3, to account for missed $r < 7\,$kpc pairs.

At $z > 0.3$, we show the fraction of
$M_B \la -20$ galaxies in close pairs with $\la 30$\,kpc separation
(their $\le 20 h^{-1}_{100}$\,kpc values) 
taken from \citet[open diamonds]{lefevre00}; we have 
adjusted the values downwards to 65\% of their original
values to account for projection within galaxy groups (as discussed above, 
as opposed to projection of random galaxies along the line of
sight, which Le F\`evre et al.\ corrected for already).
We do not show values from \citet{bundy03}, whose optical pair
statistics agree well with Le F\`evre et al.'s values but who argue
(based on near-infrared data) 
that many of the apparently luminous pairs are in fact minor mergers
which have been boosted in rest-frame $B$-band luminosity by enhanced
star formation\footnote{We address this source of concern through 
 the analysis of the stellar mass-limited sample, but find 
a roughly equal pair fraction. On one hand, many pairs of luminous galaxies
are indeed minor mergers by mass; yet, on the other hand, 
there are a large number of 
red and lower-luminosity galaxies which are missed by the luminous 
galaxy criterion which make it into a mass-limited sample.  }.
We attempt also to include the estimates of 
\citet{lin04}.  They adopt $H_0 = 70\,$km\,s$^{-1}$\,Mpc$^{-1}$
for the purposes of quoting $k$-corrected magnitudes, thus
their evolution-corrected $-21 \le M_B^e \le -19$ sample
corresponds roughly to $-22 \la M_B \la -20$, remembering 
that the evolution correction is roughly 1 magnitude per unit
redshift.  They quote their pair fractions in terms of $h_{100} = 1$
(L.\ Lin, 2006, priv.\ comm.) thus their $10 < r / h^{-1} {\rm kpc} < 30$
bin corresponds to $15 < r < 42$\,kpc adopting our $H_0$.  Since
we find that $P(r < r_f) \propto r_f$, their 27\,kpc of 
coverage for their pair fraction should be approximately 
equal to the $P(r < r_f)$ which we would calculate within 30\,kpc.  
We do, however, apply
a correction of 0.65 to their measurements, to account for 
projection at small radii (following \S \ref{proj}).
It is clear that the COMBO-17 $0.4<z<0.8$ estimate is quantitatively
consistent with these estimates, to within the combined uncertainties,
with the advantage
of robust projection correction, a volume-limited
galaxy sample, large sample size, and therefore 
highly competitive errors.  

Owing to the difficulty in extracting the properties of 
progenitors from a morphologically-classified ongoing merger, 
we elect not to compare explicitly with morphologically-derived
merger fractions from \citet{lefevre00}, \citet{conselice03}, 
\citet{cassata05}, and \citet{lotz06}.  
\citet{lotz06} compare their merger fractions to the other morphological
studies, finding consistent results to within their combined
error bars.  They further compare their results with 
Lin et al.'s and Patton et al.'s results for pair fraction 
evolution, finding overall consistency in both the inferred
zero point and redshift evolution. 

\subsubsection{The $M_* > 2.5 \times 10^{10} M_{\sun}$ sample}

To the best of our knowledge, there is only one published estimate
of the massive galaxy merger fraction at intermediate redshift, 
for $M_* > 10^{10} M_{\sun}$ 
galaxies from \citet{conselice03}.  This estimate is of limited
applicability: not only is the mass limit substantially lower, but it 
is derived from ongoing mergers, making it almost impossible 
to fairly compare with our pair-based estimate.  
Furthermore, the small sample
size weakens their constraints on the merger fraction; in particular, 
they can only place an upper limit at intermediate redshift.  
Nonetheless, their constraints are consistent with ours, to within
the combined uncertainties.  

\citet{xu04} presented an analysis of $K$-selected galaxy pairs 
taken from a combined 2MASS/2dFGRS sample \citep{cole01}.  Converting 
their results to our value of $H_0$ and stellar IMF, extrapolating
their $7.5<r<30$\,kpc results to $r<30$\,kpc, and accounting for 
the $\sim 30$\% contamination
of their pair sample with `group interlopers', we find a pair 
fraction of $1\% \pm 0.5\%$ for galaxies with 
$M_* > 2.5 \times 10^{10} M_{\sun}$, in excellent agreement with the 
SDSS determination.

\subsection{The relationship between 
close pair fraction and merger 
fraction}
\label{mrgpair}

In order to compare with galaxy formation models and 
to explore the implications of these and other 
close pair fraction determinations for galaxy merger
rate, it is important to discuss the relationship between 
galaxy close pair fraction and merger fraction 
\citep[see, e.g.,][for a recent discussion]{lotz06}.

Let us, for the sake of argument, take the case
of galaxy mass limited samples.  In this work, 
we derive the close pair fraction $f_{\rm pair} = N_{\rm gal,pair} / N_{\rm M_* > 2.5 \times 10^{10} M_{\sun}}$, where
$N_{\rm gal,pair}$ is the number of galaxies in pairs with 
${\rm M_* > 2.5 \times 10^{10} M_{\sun}}$, and 
$N_{\rm M_* > 2.5 \times 10^{10} M_{\sun}}$ is the number of massive
galaxies, per given volume.  If one wanted to relate
this close pair fraction to a merger fraction, i.e., the fraction 
of galaxies which have been created by mergers of 
galaxies in the pair sample, one notes that 
{\it i)} two galaxies in pairs merge into one merger remnant, {\it ii)}
the merger remnants are higher mass, and {\it iii)} the timescales
of being a recognizable merger remnant may differ from the close
pair timescale.  In this example above, the merger fraction most 
directly related to the above close pair fraction is:
$f_{\rm merg} = N_{\rm merg} /  N_{\rm M_* > 5 \times 10^{10} M_{\sun}}$; 
i.e., the number of newly-created merger 
remnants with $M_* > 5 \times 10^{10} M_{\sun}$
(the remnants of mergers between galaxies with ${\rm M_* > 2.5 \times 10^{10} M_{\sun}}$) divided by the number density of $M_* > 5 \times 10^{10} M_{\sun}$
galaxies.  The number of pairs should be then be related to the number
of newly-created remnants: $N_{\rm gal,pair} = 2 N_{\rm merg} \tau_{\rm pair}/\tau_{\rm merg}, $
where $\tau_{\rm pair}/\tau_{\rm merg}$
is the ratio of the timescales over which a pair enters one's close pair 
sample vs.\ the timescale over which a merger remnant is 
recognizably disturbed, and 
the factor of two accounts for the fact that a galaxy pair merges
to form a single remnant.

Let us make this example more concrete.  
\begin{eqnarray}
f_{\rm pair} & = & N_{\rm gal,pair} / N_{\rm M_* > 2.5 \times 10^{10} M_{\sun}} \\
 & \sim &  \frac{2 N_{\rm merg} \tau_{\rm pair}/\tau_{\rm merg}}{N_{\rm M_* > 2.5 \times 10^{10} M_{\sun}}} \\
 & = & \frac{2 N_{\rm merg} \tau_{\rm pair}/\tau_{\rm merg}}{\alpha N_{\rm M_* > 5 \times 10^{10} M_{\sun}}}, 
\end{eqnarray}
where $\alpha$ is the ratio in number density between galaxies with 
$M_* > 2.5 \times 10^{10} M_{\sun} $ and 
$M_* > 5 \times 10^{10} M_{\sun}$.  This ratio
$\alpha \sim 2$ in this case (as directly measured from the dataset), giving
 $f_{\rm pair} \sim f_{\rm merg} \tau_{\rm pair}/\tau_{\rm merg}$.
Thus, the fact that two galaxies in pairs merge to form only a single
remnant (entering the numerator of the fractions) is
canceled
out by the factor of two different number densities (in our particular
case) between the pair parent population and the plausible newly-created 
remnant
population (entering in the denominator 
of the merger fractions).  
Many previous analyses neglect or underestimate this 
difference in the number density of the parent population from which 
the pairs are drawn vs.\
the number density of the higher mass merger remnants; this is one
of the main contributors to our higher inferred merger rate than 
those estimated by \citet[see \S \ref{muse} for further discussion]{lin04}.
A similar argument, with similar outcome, applies
to the luminous galaxy sample.  We will use this argument in what follows.

\subsection{Comparison with galaxy formation models}

In this section, we compare model major merger remnant fractions with our
measurements of close pair fraction 
\citep[see][for a detailed discussion of model insights into the meaning and evolution of close pair fraction]{berrier}.  In what follows, 
we adopt $\tau_{\rm merg}$ = $\tau_{\rm pair}$ = 0.4\,Gyr, 
following the estimated pair timescale calculated in 
the next section.  Thus, $f_{\rm merg} \sim f_{\rm pair}$, 
as the timescales are defined to be equal, as long 
as we choose to explore the fraction of galaxies recently created in 
galaxy mergers 
with $M_* > 5 \times 10^{10} M_{\sun}$ and/or 
$M_B < -20.75$.  

In both panels of Fig.\ \ref{fig:frac2}, 
we show major merger fractions taken from an updated
version of the Somerville et al.\ semi-analytic galaxy formation model
\citep[see][for a description of the basic model ingredients]{sp,spf}.
The model includes standard prescriptions for gas cooling, feedback,
and dust extinction and star formation. Quiescent star formation is
parameterized as in \citet{delucia04}, and bursts of star formation
are triggered by major and minor mergers, based on results from
hydrodynamic simulations of merging galaxies \citep{tjthesis}.
Feedback from AGN is not included in this model, although we found
that merger rates calculated from a model including AGN feedback were
similar to those presented here.  The model reproduces reasonably well
the evolution of the luminosity and stellar mass function of galaxies
in the interval $0<z<1$.  This particular model uses Monte Carlo
realizations of dark matter merger histories based on the analytic
Extended Press-Schechter formalism, supplemented with standard
prescriptions for dynamical friction, and therefore we lack detailed
information about the spatial location of galaxies within their dark
matter halos. As a result, we cannot directly compute close pair
fractions from these simulations; rather, we compute the fraction of
galaxies which have undergone a recent merger and compare this with
the observational estimates.

In the right-hand panel of Fig.\ \ref{fig:frac2}, we
show also the close pair fraction inferred from 
the merger rates of massive galaxies from a SPH 
galaxy formation model 
\citep[dotted line]{maller05}, under the same timescale
and merger vs.\ pair assumptions as adopted above.
  \citet{maller05}
present the fraction of mergers per Gyr for galaxies with 
$M_* > 6.4 \times 10^{10} M_{\sun}$ with mass ratios less than 2:1
in their Fig.\ 3: they argue that this is equivalent
to a mass cut of $\sim 2.5 \times 10^{10} M_{\sun}$, as their 
model dramatically over-produces stellar mass by a factor of 
2.75, and as a stop-gap measure they suggest division of the
mass cut by that factor when comparing with data.  This mass
limit is different from the merger remnant mass limit of 
$M_* > 5 \times 10^{10} M_{\sun}$; accordingly, we treat 
this comparison as more qualitative than quantitative.
The merger rates are increased
by a factor of 1.7 to transform from 2:1 to a 3:1 
threshold (following their Fig.\ 6).   These merger
fractions are somewhat higher than observed, and show 
a steeper redshift dependence than the Somerville et al.\
model estimates.  

Recently, \citet{berrier} discussed the evolution of the 
close pair fraction of luminous galaxies, finding overall
consistency at the factor of two level with the \citet{lin04}
measurements; given that our measurements are consistent 
with the measured pair fractions from \citet{lin04}, we would
expect that the model close pair fraction from \citet{berrier}
would be reasonably consistent with both our data and the Somerville et al.\ 
model.  

While important
discrepancies remain between different models, different datasets, and 
between the data and models, it is nonetheless obvious that there is
an overall qualitative 
consistency between our best attempts at observationally
constraining the merger/close pair fraction of 
luminous and massive galaxies and our present 
understanding of galaxy formation and assembly
in a $\Lambda$CDM universe. 
Given that the 
ongoing assembly of massive galaxies is a key 
(and unavoidable) feature of galaxy evolution 
in such a cosmology, it is encouraging that 
there is a decent qualitative agreement between the 
models and data at this stage. 

\subsection{Musings on the merger rate of galaxies} \label{muse}

In order to convert close pair fractions into merger
rates, a timescale over which a close pair of nearly-equal
mass galaxies will merge is required.  The estimation of
such timescales far from straightforward and is the topic
of much ongoing work: the mix of orbital parameters
will lead to a distribution of timescales, and the effects
of e.g., dynamical friction and fly-bys in dense environments are
poorly-understood.
Here, for illustrative purposes, we take a highly simplistic approach and
assume that the merger of two nearly-equal mass galaxies 
takes roughly one orbital timescale 
$t_{\rm orb} \sim 2 \pi r / 1.4 \sigma \sim 4 r / \sigma$,
noting that the circular velocity of a galaxy is $\sim 1.4 \sigma$,
where $\sigma$ is the velocity dispersion of the galaxy in question.
This estimate is rather similar to those presented
by, e.g., \citet{patton02} or \citet{lin04}, and compares favorably to 
timescales derived from the \citet{naab06} dry galaxy merger
simulations.  We adopt a typical velocity dispersion of a luminous/massive
galaxy of $\sim 150$\,kms$^{-1}$, and we adopt the typical 
radius of a galaxy in the $r_f < 30$\,kpc sample of $r \sim 15$\,kpc.
On this basis, we estimate a merger timescale of $\sim 0.4$\,Gyr; 
uncertainties in this timescale
are at least a factor of two\footnote{It is worth 
noting that if dynamical friction timescale 
arguments were used, one would derive timescales
for $r \sim 15$\,kpc of $\sim 0.3$\,Gyr, following
the discussion of dynamical friction in \citet{binney}.  
While it is true that the Chandresekar formulation 
of dynamical friction should not strictly apply
in the case of a merger between two massive galaxies, it is 
nonetheless encouraging that the orbital timescale
and dynamical timescale arguments yield approximately equal timescales.}. 
Recalling that $f_{\rm merg} \sim f_{\rm pair}$ when 
equal timescales are chosen (from \S \ref{mrgpair}) and 
when the pair fraction from lower luminosity/mass samples is related
to the merger fraction of the more massive remnant galaxies, 
the merger rate per Gyr is $P(r<30\,{\rm kpc}) / 0.4$.

Given this rough estimate for merger rate, one can derive the 
rate of creation of galaxies with $M_* > 5 \times 10^{10} M_{\sun}$ 
through major mergers using 
the estimated close pair fractions of 
galaxies with $M_* > 2.5 \times 10^{10} M_{\sun}$ at $z \sim 0.1$ from SDSS
coupled with the $z\sim 0.6$ determination from COMBO-17.
A power-law fit to these data points yields: $\log f \sim -2.1\pm0.2 + 
[3.9\pm0.5 \log (1+z)]$, where $f$ is the close pair fraction.  Using
that the creation rate of merger 
remnants with $M_* > 5 \times 10^{10} M_{\sun}$  is  $\sim f / 0.4$, and 
integrating as a function of cosmic time, one derives 
an average of $\sim 0.5\,(0.7)$ galaxy mergers per present-day 
$M_* > 5 \times 10^{10} M_{\sun}$ galaxy since $z = 0.8\,(1)$.
The corresponding numbers for galaxies with $M_B < -20.75$, using all 
available observational determinations and weighting by the uncertainties, 
are: $\log f \sim -1.9\pm0.2 + 
[2.6\pm0.8 \log (1+z)]$, and $\sim 0.5\,(0.6)$   mergers
since $z = 0.8\,(1)$.  This determination is very similar
to the recent determination of \citet[0.3-0.7 mergers per 
luminous galaxy]{lotz06}\footnote{Differences 
between this determination and that of \citet{lin04} have a 
factor of 1.3 contribution from 
inclusion of our data and that of the 2dFGRS and \citet{lefevre00}, 
a factor of 2 difference following \S \ref{mrgpair}, a factor of 1.2
difference in redshift integration method, a factor of 1.25 
difference in timescale (our 0.4\,Gyr vs.\ their 0.5\,Gyr), and 
finally a difference of 1.3 from the fraction of their pairs deemed to be 
real (from our sample we estimated 65\% of the pairs are physical; they argued
50\% following the low redshift analysis of Patton et al.): 
in all, a factor of five difference in inferred remnant density would 
be expected, and indeed our estimate is five times larger than theirs.}.
While it is clear that improvements
in both the observational determination of close pair
fraction evolution with redshift and work towards robust
estimates of merger timescale are of critical importance, 
this analysis strongly suggests that 
galaxy mergers between luminous/massive
galaxies are a significant feature of the last eight billion years
of galaxy evolution.

\acknowledgements
The referee is thanked for their constructive, thought-provoking and
and thorough critique of the draft.
It is a pleasure to thank David Hogg and Morad Masjedi for their 
important contributions to building up the methodology 
used in this paper; Frank van den Bosch and Romeel Dav\'e are 
thanked for useful discussions. 
Elizabeth Barton is thanked for her comments on an early draft of the paper.
E.\ F.\ B.\ and S.\ P.\ were supported by the European Community's Human
Potential Program under contract HPRN-CT-2002-00316 (SISCO); E.\ F.\ B.\
is currently supported by  
the DFG's Emmy Noether Program.
C.\ W.\ was supported by a PPARC Advanced Fellowship.

\end{document}